\documentclass[amsmath,amssymb,aps,twocolumn,superscriptaddress]{revtex4-2}

\usepackage{graphicx}
\usepackage[dvipsnames]{xcolor}
\usepackage{hyperref}
\hypersetup{
  colorlinks   = true, 
  urlcolor     = Blue, 
  linkcolor    = Blue, 
  citecolor   = Blue 
}
\usepackage[mathlines]{lineno}
\modulolinenumbers[5] 

\begin{document}


\title{First observation of neutral bremsstrahlung electroluminescence in liquid argon}


\author{A. Bondar}
\author{A. Buzulutskov}
\thanks{Deceased on September 26th, 2023}

\author{E. Frolov}
\email[]{geffdroid@gmail.com}

\author{E. Borisova}

\author{V. Nosov}

\author{V. Oleynikov}

\author{A. Sokolov}
\affiliation{Budker Institute of Nuclear Physics SB RAS, Lavrentiev avenue 11, 630090 Novosibirsk, Russia}
\affiliation{Novosibirsk State University, Pirogova street 2, 630090 Novosibirsk, Russia}


\date{\today}

\begin{abstract}
A recent discovery of additional mechanism of electroluminescence (EL) in noble gases due to the neutral bremsstrahlung (NBrS) effect led to a prediction that NBrS EL should be present in noble liquids as well. A theoretical model of NBrS EL in noble liquids was developed accordingly in the frameworks of Cohen-Lekner and Atrazhev. In this work, we confirm this prediction: for the first time, visible-range EL has been observed in liquid argon at electric fields reaching 90~kV/cm, using Gas Electron Multiplier (GEM) and Thick GEM (THGEM) structures. Absolute light yields of the EL were measured and found to be in excellent agreement with the theory, provided that the momentum-transfer cross section of electron scattering is used for calculation of NBrS cross section (instead of the energy-transfer one).
\end{abstract}


\maketitle



\section{\label{intro}Introduction}
Electroluminescence (EL) is an optical and electrical phenomenon in which a material emits light in response to electric current or electric field. Of paramount importance is EL in noble gases, as it is a key physical process used in two-phase (liquid-gas) detectors for dark matter searches and neutrino detection experiments~\cite{Akimov21,Aprile18,Agnes18b,LUX16}. In two-phase detectors both a prompt primary scintillation signal (S1) and a delayed primary ionization signal (S2) are measured, the latter being recorded in the gas phase using the EL effect. 

According to modern concepts~\cite{Akimov21,Buzulutskov20}, there are three mechanisms responsible for EL in noble gases: that of excimer emission in the vacuum ultraviolet (VUV), that of neutral bremsstrahlung (NBrS) emission in the UV, visible and near infrared (NIR) range and that of emission due to atomic transitions in the NIR. These three mechanisms are referred to as excimer (ordinary) EL, NBrS EL and atomic EL, respectively. Let us briefly describe the first two using the example of argon.

Excimer EL is due to noble gas excimers in a singlet (Ar$^{*}_{2}(^{1}\Sigma^{+}_{u})$) or triplet (Ar$^{*}_{2}(^{3}\Sigma^{+}_{u})$) state emitting photons in the VUV (around 128 nm). The excimers are produced in three-body atomic collisions of the lowest excited atomic states, of Ar$^*(3p^54s)$ configuration, which in turn are produced by drifting electrons in electron-atom collisions~\cite{Akimov21,Chepel13,Buzulutskov20,Buzulutskov17,Oliveira11}. 
It has a threshold in reduced electric field (E/N) of about 4~Td (1~Td$=10^{-17}$~V~cm$^2$)~\cite{Oliveira13,Buzulutskov22}.

NBrS EL is due to bremsstrahlung of slow drifting electrons scattered on neutral atoms~\cite{Buzulutskov18,Borisova21,Borisova22,Aoyama22,Henriques22,Milstein22,Milstein23}. NBrS EL has no threshold in electric field, takes place in the visible and NIR range suitable for direct readout by conventional photodetectors such as photomultiplier tubes (PMTs) and silicon photomultipliers (SiPMs), but has much lower photon yield compared to that of excimer EL~\cite{Borisova21}. 

While EL in noble gases is well understood~\cite{Buzulutskov20,Oliveira11,Oliveira13,Buzulutskov18,Borisova21}, little is known about EL in noble liquids. Firstly, until recently, there was no complete theory of EL in noble liquids. Secondly, excimer EL was investigated in detail in only three experiments conducted in liquid xenon~\cite{Masuda79,Aprile14b,Qi23} and was observed above the electric field threshold of 400~kV/cm. There were other observations of EL in liquid xenon~\cite{Schussler00,Ye14}, but neither the EL threshold nor the yield values were obtained there. In order to obtain such high electric fields it is necessary to use thin wires, needles or hole-like structures such as Gas Electron Multipliers (GEM) or Thick GEMs (THGEM)~\cite{Sauli16,Bressler23}.

In liquid argon, excimer EL has never been observed, presumably due to its very high threshold expected from theoretical estimations, exceeding 800~kV/cm \cite{Borisova22,Stewart10}. It should be remarked that proportional EL claimed to be observed in liquid argon in~\cite{Lightfoot09}, at much lower electric fields (of about 60~kV/cm) and using THGEM, apparently was not produced in liquid argon, but rather in gas bubbles associated with THGEM holes, similarly to what happens in Liquid Hole Multipliers~\cite{Erdal20} (see discussion in~\cite{Borisova22}). 

As predicted in~\cite{Buzulutskov18}, by its universal nature NBrS EL should be present in noble liquids at electric fields much lower than those needed for excimer EL. A theoretical model of NBrS EL in noble liquids has been recently developed accordingly~\cite{Borisova22,Henriques22}, the electron energy and transport parameters being obtained in the framework of Cohen-Lekner~\cite{Cohen67} and Atrazhev~\cite{Atrazhev85}. The observation of NBrS EL in noble liquids, in addition to the obvious interest in it as a new physical effect, would also prompt the search for new readout schemes in single-phase noble liquid detectors of improved performance.

In this work, to resolve the issues described above, true visible-range EL in liquid argon has been studied for the first time, using both THGEM and GEM structures. We will show that the absolute light yields of EL observed in experiment are in excellent agreement with those predicted by the NBrS EL theory appropriately modified compared to~\cite{Borisova22}.

\section{\label{SetupSection}Experimental setup}
\begin{figure}[!t]
\centering
\includegraphics[width=0.53\linewidth]{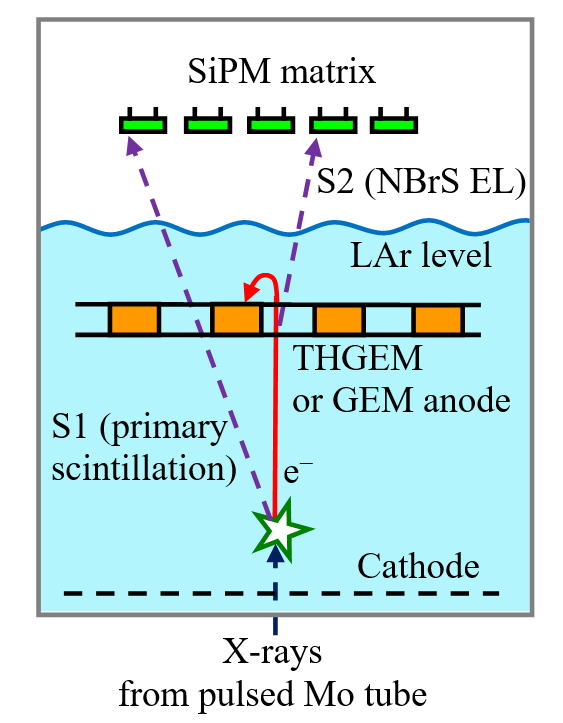}
\caption{\label{fig:setup_scheme}Conceptual illustration of experimental setup used in this work (not to scale).}
\end{figure}

The experimental setup used in this work is described elsewhere~\cite{Buzulutskov22,Bondar22} and thus we recount essential points only. A few minor modifications described below were made to allow for the study of visible EL in liquid argon using detector configuration schematically depicted in Fig.~\ref{fig:setup_scheme}.

The detector was a single-phase liquid time projection chamber (TPC), composed of a drift region and a THGEM (GEM) anode where EL took place. It was filled with 3 liters of purified liquid argon ($<$4~ppb of O$_2$ and $<$1~ppm of N$_2$ impurity) and operated at a pressure of 1.00~atm, temperature of 87.3~K and atomic density of the liquid of 2.10$\cdot$10$^{22}$~cm$^{-3}$~\cite{Fastovsky72,Stewart89}. 

A THGEM or GEM plate immersed in the liquid was used to study EL produced in its holes by applying the voltage across it, thus creating the region of high electric field inside the holes. The light from primary scintillation (prompt S1 signal) as well as the light from the holes (delayed S2 signal) was directly recorded by a 5$\times$5 matrix of SiPMs of 13360-6050PE type~\cite{Hamamatsu} facing the THGEM (GEM) plate and sensitive in the visible and NIR range. The THGEM had a dielectric thickness of 0.4~mm, hole pitch of 0.9~mm, rim of 0.1~mm, hole diameter of 0.5~mm and copper thickness of 0.03~mm. The GEM had the following parameters: dielectric thickness of 50~$\mu$m, hole pitch of 140~$\mu$m, zero rim, copper thickness of 5~$\mu$m, outer hole diameter of 70~$\mu$m and inner hole diameter of 50~$\mu$m (biconical hole design).

Pulsed X-rays from an X-ray tube with molybdenum anode~\cite{Bondar16}, with the average energy deposited in liquid argon of 25~keV, were used as an ionization source. The X-ray tube also provided the external trigger. In order to measure the absolute EL yield, it is necessary to know the number of electrons in a pulse that reach the THGEM (GEM) holes. To this end, the charge arriving to THGEM (GEM) was recorded directly in special calibration runs using a charge-sensitive preamplifier~\cite{Bondar22}. The preamplifier was calibrated in a standard procedure~\cite{James20} using generated pulses sent to the input of the circuit via an injection capacitor.

\section{\label{Theory}Theory}

In this work, we compare the measurements of visible-range EL in liquid argon with the theoretical model of NBrS EL developed in~\cite{Borisova22}. The theory is based on the compact formula approximating the cross section of NBrS photon emission via cross section of electron elastic scattering on atom~\cite{Buzulutskov18,Henriques22,Firsov60,Kasyanov65,Dalgarno66,Biberman67,Park00}:

\begin{eqnarray}
\frac{d\sigma}{d\nu} = \frac{8}{3} \frac{r_e}{c} \frac{1}{h\nu}&& \left(\frac{\varepsilon-h\nu}{\varepsilon}\right)^{1/2} \nonumber\\ &&\times [(\varepsilon-h\nu)\sigma_{el}(\varepsilon) + \varepsilon\sigma_{el}(\varepsilon - h\nu)]\;, \label{NBrS-XS}
\end{eqnarray}
where $h\nu$ is the photon energy, $r_e = e^2/mc^2$ is the classical electron radius, $c$ is the speed of light, $\varepsilon$ is the energy of an incident electron and $\sigma_{el}(\varepsilon)$ is the cross section of its elastic scattering on atom. Note that there are two elastic cross sections that can be determined empirically: a total elastic one and a transport (momentum transfer) one, the latter describing the transport parameters such as electron drift velocity and diffusion coefficients. It was proposed in~\cite{Borisova22} that within the Cohen-Lekner~\cite{Cohen67} and Atrazhev~\cite{Atrazhev85} approach this formula, initially derived for gas, can also be applied for liquid.

The absolute EL yield, $Y_{EL}$, is defined as the number of photons produced per unit drift path and per drifting electron. To compare results at different medium densities and temperatures, reduced EL yield, $Y_{EL}/N$, is used instead, where $N$ is liquid atomic density. For NBrS EL it can be described by the following equation, using the electron energy distribution function $f(\varepsilon)$ normalized to unity \cite{Buzulutskov18}:
\begin{equation}
\label{NBrS-Y}
\frac{Y_{\mathrm{EL}}}{N} = \int_{\lambda_{1}}^{\lambda_{2}}\!\!\! \int_{h\nu}^{\infty}\!\!\frac{\sqrt{2\varepsilon/m}}{v_d} \frac{d\sigma}{d\nu}\frac{d\nu}{d\lambda}\, f(\varepsilon)\, d\varepsilon\, d\lambda \;, 
\end{equation}
where $v_d$ is the electron drift velocity and $\lambda_1$--$\lambda_2$ is the wavelength region of interest. In this work as well as in our previous ones the latter is limited to the wavelength region of 0-1000~nm.

To obtain the electron energy distribution function, needed to calculate the EL yield in liquid argon, the Cohen-Lekner~\cite{Cohen67} and Atrazhev~\cite{Atrazhev85} approach was used here, similarly to that of~\cite{Borisova22}, in which the electron transport through the liquid is considered as a sequence of single scatterings on an effective potential. In particular, the electron scattering cross section can be used in the liquid in a way similar to that of the gas. An important concept for this approach is a distinction between the energy transfer scattering, which changes the electron energy, and that of momentum transfer, which only changes the direction of the electron velocity. Both processes are assigned separate cross sections~\cite{Stewart10,Cohen67,Atrazhev85} differing by a structure factor which appears when the coherence of scattering is taken into an account. Similarly to the total elastic and transport cross sections in gases, the energy transfer and momentum transfer cross sections in liquids describe the same process of electron scattering and are only different integrals of an effective differential cross section. Also similarly to gases, both cross section are determined empirically and those given in~\cite{Atrazhev85} are used here, since then the theory describes well the experimental data on the electron drift velocity in liquid argon~\cite{Borisova22,Atrazhev85}. It should be noted that these two cross sections are both transport ones and therefore are both used for calculating electron transport parameters.

\begin{figure}[!t]
\includegraphics[width=1.0\linewidth]{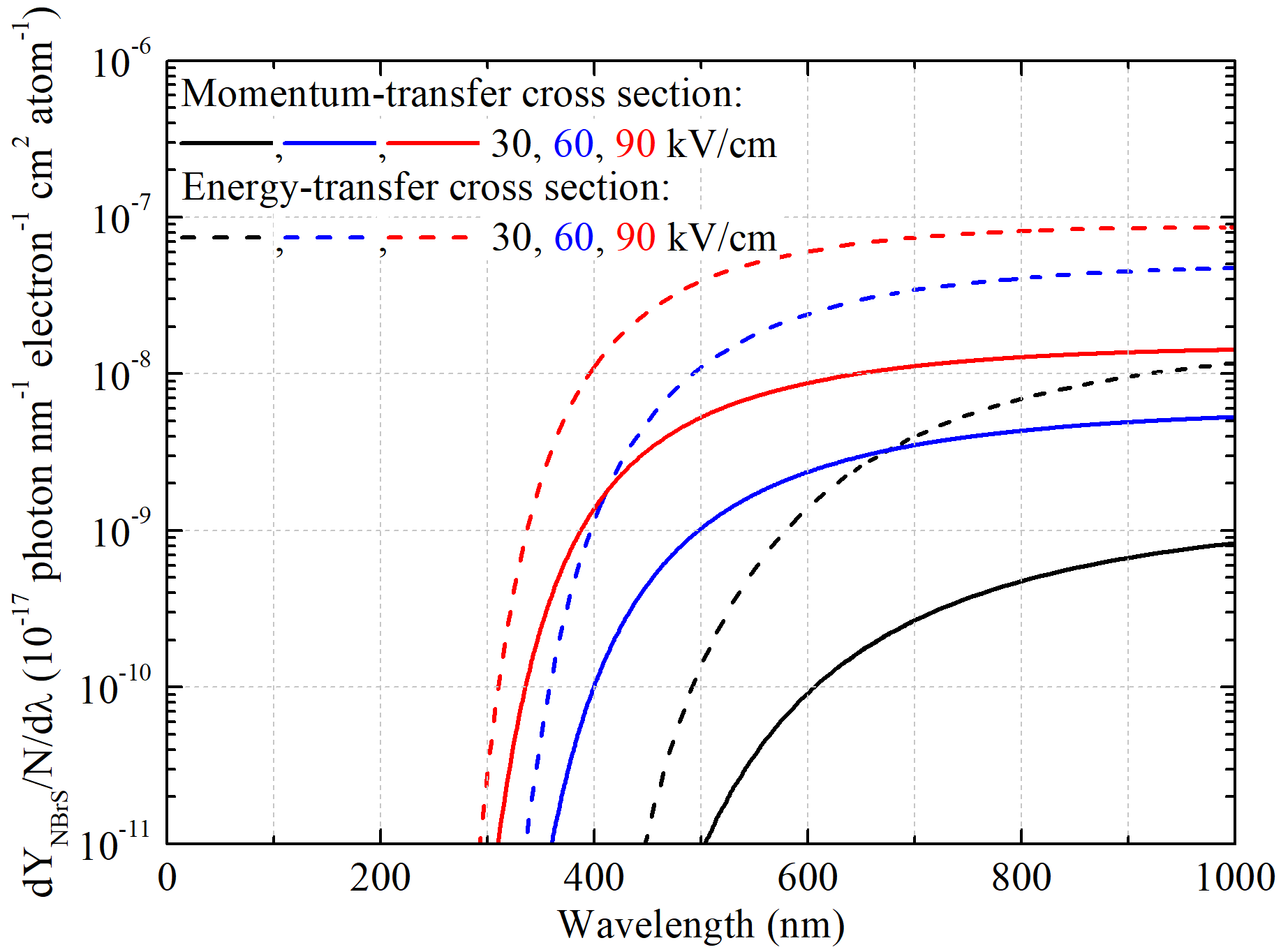}
\caption{\label{fig:spectra}Theoretical spectra of the reduced EL yield for NBrS EL in liquid argon at different electric fields. The results are shown for the energy-transfer and momentum-transfer cross sections being used in Eq.~(\ref{NBrS-XS}), both taken from~\cite{Atrazhev85}.}
\end{figure}

The wavelength spectra of the reduced EL yield for NBrS EL in liquid argon, obtained from Eq.~(\ref{NBrS-Y}) by taking its derivative with respect to~$\lambda$, are shown in Fig.~\ref{fig:spectra} at different electric fields. One can see that they are rather flat and do not differ much in shape from those of gaseous argon~\cite{Borisova21}.

It should be remarked that there is an ambiguity about what cross section of elastic scattering, $\sigma_{el}(\varepsilon)$, should appear in Eq.~(\ref{NBrS-XS}): some theoretical derivations show that it should be the total elastic cross section~\cite{Firsov60,Kasyanov65,Dalgarno66,Biberman67,Kasyanov78}, while others show that it should be the transport one~\cite{Milstein22,Dalgarno66,Biberman67,Park00}. The former derivations use asymptotic electron wave functions and~\cite{Kasyanov65} in particular yields Eq.~(\ref{NBrS-XS}) if one assumes that the electron scattering is dominated by a single partial wave. The latter ones derive the NBrS cross section in a limit of small photon energies $h\nu~\rightarrow~0$, suggesting that it is the transport cross section that should be used in Eq.~(\ref{NBrS-XS}). In gases these two approaches do not result in significantly different NBrS yields~\cite{Buzulutskov18,Henriques22,Amedo22} and are both in good agreement with experiment.

Similarly to gases, in liquids it is not clear which of the two available cross sections of electron scattering on the effective potential should be used in Eq.~(\ref{NBrS-XS}): the energy-transfer one or the momentum-transfer one. In contrast to the gases, the results on the NBrS emission using different cross section can differ by more than an order of magnitude, depending on the electric field. One can see this in Fig.~\ref{fig:spectra} showing the NBrS EL spectra for when either of the cross sections is used in Eq.~(\ref{NBrS-XS}). Experimental measurements of the absolute EL yields should determine the best approach and whether the approximate formula Eq.~(\ref{NBrS-XS}) for the NBrS emission initially derived for gases works for liquids as well.

In practice one needs to know the absolute light yield of EL for a given device, defined as the total number of EL photons (with $\lambda$$\le$1000 nm) produced in the THGEM or GEM holes to a full solid angle (4$\pi$) per drifting electron, at a given electric field: $Y_{THGEM}$ or $Y_{GEM}$. This quantity was calculated by Monte Carlo simulation of electron drift through the THGEM (GEM) holes accompanied by the NBrS photon emission, using Eq.~(\ref{NBrS-Y}) and its spectral derivative. 

As a first step, precise electric field map in the THGEM and GEM was calculated using Gmsh~\cite{Gmsh,Geuzaine2009} and Elmer~\cite{Elmer} open-source programs in the same way as in~\cite{Bondar19}. Then, using the field map and electron transport parameters such as electron drift velocity and diffusion coefficients (also obtained following Atrazhev formalism~\cite{Borisova22,Atrazhev85}), the electron drift through the THGEM (GEM) was simulated using the TPC geometrical model built with Geant4 library~\cite{Geant4a,Geant4b,Geant4c}. The drift of electrons was implemented using algorithms from Garfield++ library~\cite{GarfieldPP}. The starting positions of the electrons were defined by geometry of X-ray source used in the experiment. Finally, using electron drift tracks comprised of multiple small steps, NBrS photons were generated along each step at a given electric field, using known field dependence of the NBrS spectra derived from Eq.~(\ref{NBrS-Y}).

Since in the experiment the number of photoelectrons on the SiPM matrix per drifting electron is actually measured, the theory should be able to convert it to $Y_{THGEM}$ ($Y_{GEM}$). For this reason, a standard Geant4 code was used to simulate propagation and detection of photons from the THGEM (GEM) holes to the SiPM matrix according to optical parameters of the detector materials. Then the photon detection efficiency (PDE) of the SiPMs, derived from \cite{Hamamatsu,Nuruyev20,Otte17}, was used to convert the photon number to photoelectron number.

From the simulation, the light collection efficiency on the SiPM matrix (LCE) from the THGEM (GEM) holes as well as SiPM PDE averaged over the detected NBrS photons were obtained. The former was found to be independent of the voltage across the THGEM (GEM). A systematic uncertainty of the LCE is estimated to be 20\% and is mainly due to limited knowledge of optical properties of materials in the detector.

The SiPM PDE averaged over detected NBrS photons depended on the electric field in the THGEM (GEM) holes: it smoothly increased from 8.6\% at 30~kV/cm to 13.4\% at 60~kV/cm and 15.8\% at 90~kV/cm. The relative uncertainty of the average PDE is estimated as 5.2\% and is due to the uncertainty of absolute PDE measurements~\cite{Hamamatsu,Nagai19,Siyang19,Nuruyev20} and the uncertainty of bias voltage at room temperature~\cite{Hamamatsu,Siyang19,Nuruyev20} which is required to obtain SiPM PDE at cryogenic temperature~\cite{Otte17}. Both the LCE and average PDE also almost do not depend on whether the energy-transfer or momentum-transfer cross section is used in NBrS calculation.

It was found during the simulations, that despite the field non-uniformity in the holes, a parallel-plate approximation of the THGEM (GEM) hole can be successfully applied since it provided the light yield close to that of exact simulation (within 30\%). In this approximation, the hole is approximated as a parallel-plate counter with the uniform electric field equal to that in the hole center and with the thickness equal to the inter-electrode distance (dielectric thickness).
Notably, such approximation worked well in the past to calculate ionization coefficients in noble gases at low temperatures using GEMs~\cite{Buzulutskov05}. Because of this, the following results are shown as a function of the electric field at the THGEM (GEM) hole center (E$_{\mathrm{hole}}$). Note that according to calculations, the electric field at the hole center is 57\% and 68\% of the ``nominal'' field, defined as the applied voltage divided by the dielectric thickness, for the THGEM and GEM respectively.

\section{\label{Results_x-ray}Experimental results}
\begin{figure}[!t]
\includegraphics[width=1.0\linewidth]{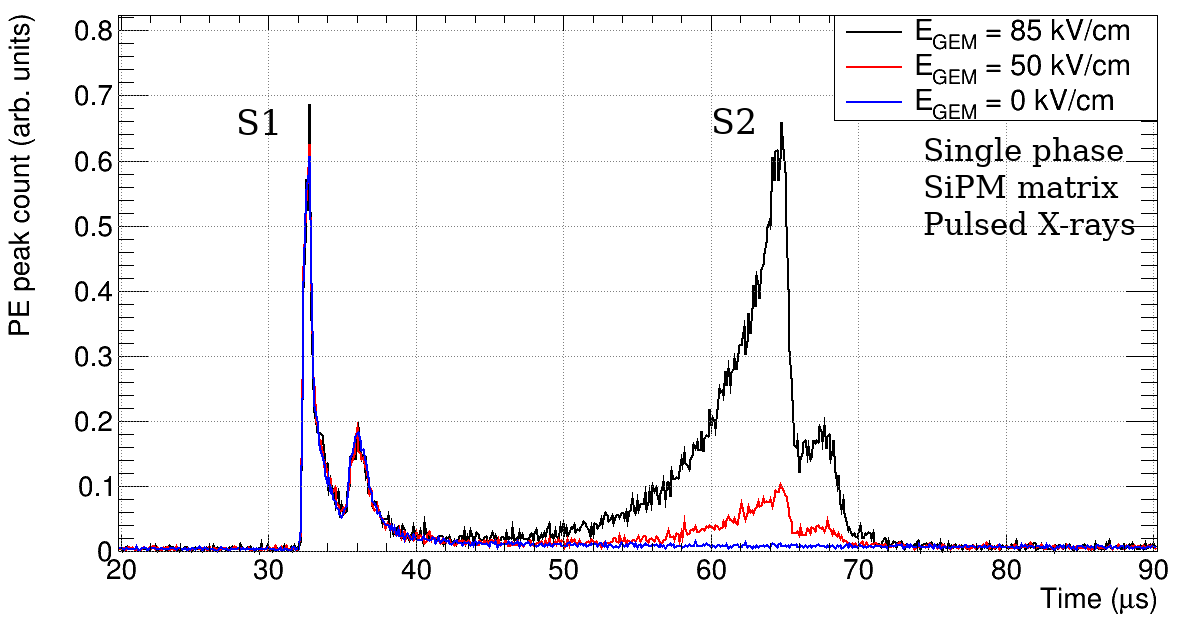}
\caption{\label{fig:X-ray_shape}Average signal pulse-shape from the SiPM matrix in liquid argon TPC with GEM anode at different voltages across it, corresponding to 0, 50 and 85~kV/cm electric field at the GEM hole center. S1 (primary scintillation) and S2 (electroluminescence in the GEM holes) signals are seen. The pulse-height is expressed in the relative number of photoelectron counts. 
S1 and S2 signals have double-peak structure due to characteristic double-pulse structure of the X-ray tube itself~\cite{Bondar12a}.}
\end{figure}

Using an intensive X-ray source, which produces about 1.4$\cdot$10$^{5}$ electrons in a pulse that escape recombination, allows for detailed study of relatively weak visible-range EL in liquid argon. The examples of average signal pulse-shape from the SiPM matrix in liquid argon TPC obtained with GEM anode are shown in Fig.~\ref{fig:X-ray_shape}. At zero voltage across the GEM (blue line), only S1 signal is seen provided by primary scintillation in the visible range~\cite{Bondar22}.

\begin{figure}[!t]
\includegraphics[width=1.0\linewidth]{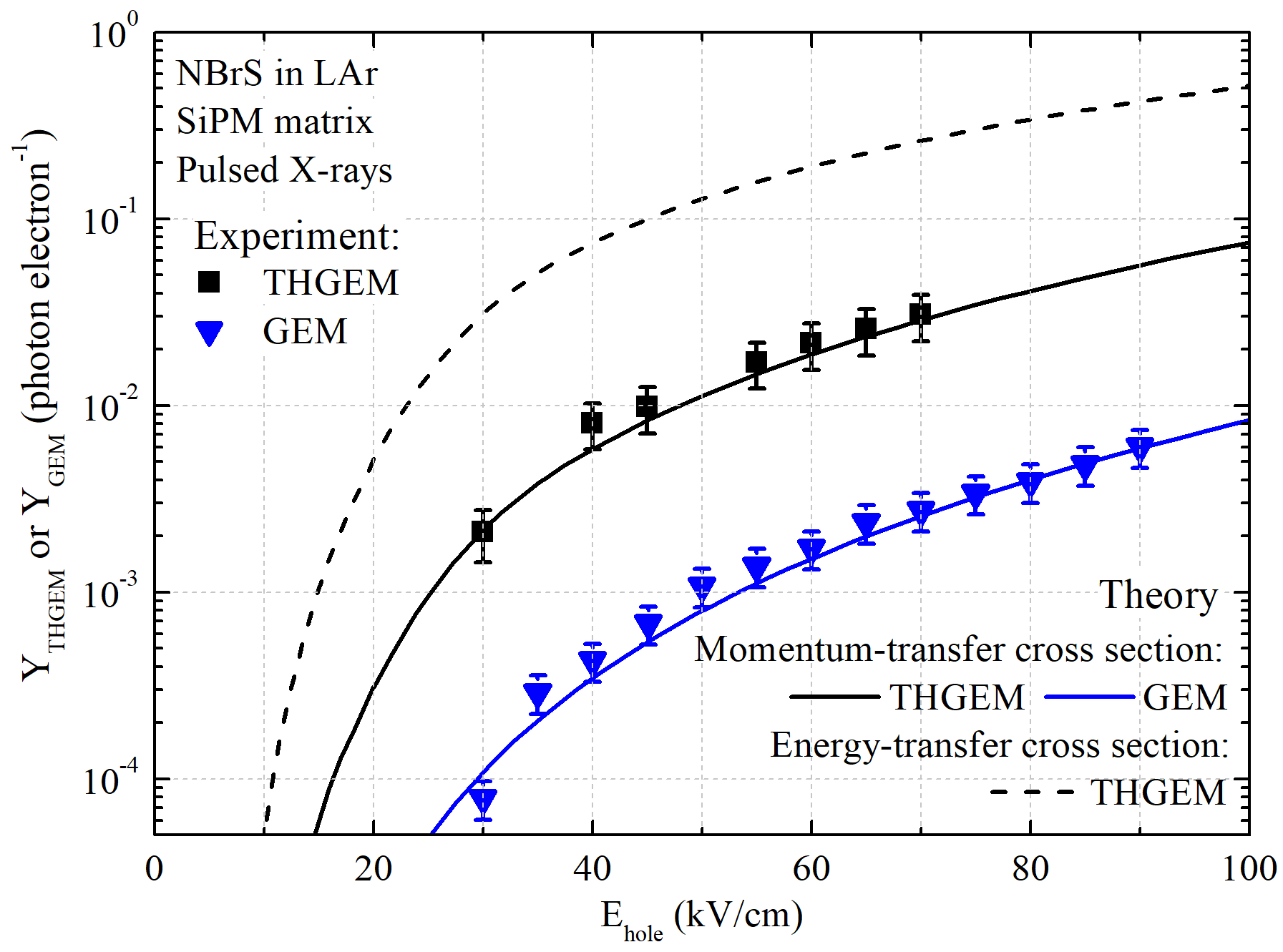}
\caption{\label{fig:X-ray_Ngamma}Absolute light yield of NBrS EL in liquid argon produced in the THGEM and GEM and expressed in photons (at $\lambda$$\le$1000 nm) per drifting electron as a function of the electric field in the hole center. Theoretical predictions for both the THGEM and the GEM when the momentum-transfer cross section is used in Eq.~(\ref{NBrS-XS}) are shown by solid lines. For comparison, the prediction for the THGEM when using the energy-transfer cross section is also shown by dashed line.}
\end{figure}

As the voltage across the GEM or THGEM increases, the S2 (electroluminescence) signal appears and increases with the electric field, without any specific threshold in contrast to that of excimer EL. It should be noted that S1 and S2 signals in Fig.~\ref{fig:X-ray_shape} have double-peak structure due to characteristic time structure of the X-ray tube itself which has two prominent peaks and a near indistinguishable third one~\cite{Bondar12a}. The front edge of S2 signal is consistent with ionization distribution along the drift electric field having 0.7--0.9~cm spread, given that the energy of X-rays arriving at drift region varies from around 11 to 40~keV~\cite{Bondar12a}.

The absolute light yield of EL produced in the THGEM or GEM ($Y_{THGEM}$ or $Y_{GEM}$) is obtained from the S2 pulse area and is expressed in photons (at $\lambda$$\le$1000~nm) per drifting electron, as defined in the previous section~\ref{Theory}. It is shown in Fig.~\ref{fig:X-ray_Ngamma} for both the THGEM and the GEM as a function of the electric field in the hole center. It should be remarked that the experimental yields shown in the figure use theoretical values of the LCE and appropriately averaged PDE to obtain the total number of photons from the number of photoelectrons recorded.

The lines in the figure show theoretical predictions obtained using Atrazhev cross sections of electron scattering~\cite{Atrazhev85} according to the procedure described in section~\ref{Theory}. The solid lines show theoretical results when the momentum-transfer cross section is used in Eq.~(\ref{NBrS-XS}), while the dashed line shows those for the energy-transfer cross section.  

As can be seen, the experimental data for both the THGEM and the GEM are in excellent agreement with the theory when the momentum-transfer cross section is used, despite an order of magnitude difference between them, while using the energy-transfer cross section gives a strongly overestimated (by an order of magnitude) yield. In other words, using the energy-transfer cross section in Eq.~(\ref{NBrS-XS}) is strongly rejected by our data.

The maximum electric fields reached in the experiment were limited by discharges. This resulted in that the maximum absolute light yield of NBrS EL in liquid argon produced in the 0.4~mm thick THGEM was about $Y_{THGEM}=3\cdot 10^{-2}$ photons per drifting electron, at the maximum electric field of 70~kV/cm at the THGEM hole center corresponding to about 4.9~kV across it. For the GEM, the light yield is an order of magnitude lower due to an order of magnitude smaller thickness.

It should be noted here that expected decrease of the light yield when replacing THGEM with GEM indicates that observed EL is not produced in gas bubbles under the plates in contrast to the previous contaminated experiments~\cite{Lightfoot09,Erdal20}. Due to EL signal being very weak it is also very unlikely that there were bubbles in THGEM (GEM) holes. Another confirmation was obtained by comparing the EL yield at 1.0 and 1.5~atm pressure in the detector since it was demonstrated in~\cite{Kim04,Erdal15} that formation of bubbles is a pressure-dependent process. We observed no significant difference in light yield at these pressures meaning that the presence of bubbles in our system is very unlikely.

There are two main sources of experimental errors shown in the Fig.~\ref{fig:X-ray_Ngamma}. The first one is the systematic uncertainty of light collection efficiency and average PDE discussed in section~\ref{Theory}. The second one is the systematic uncertainty of measuring charge reaching the THGEM (GEM) which amounted to 13\%. This results in a total 25\% systematic uncertainty. In comparison, a relative statistical error varies from about 6\% at low fields to about 2\% at high ones.

Regarding the theoretical model shown by lines in the Fig.~\ref{fig:X-ray_Ngamma}, its main uncertainty is due to uncertainty of transport cross sections which are used both in Eq.~\ref{NBrS-XS} and to obtain electron energy distributions for Eq.~\ref{NBrS-Y}. This uncertainty is driven by experimental uncertainties of liquid argon transport parameters. It is estimated that resulting light yield has an uncertainty of a factor of about 3.

\section{Conclusions}

In this work, visible-range electroluminescence (EL) in liquid argon has been observed for the first time using both THGEM and GEM structures with SiPM-matrix optical readout and pulsed X-ray source. Its absolute yield was measured for electric fields varying from 30 to 90~kV/cm. 

The observed EL can be fully explained and quantitatively described by the effect of bremsstrahlung of slow drifting electrons scattered on neutral atoms (neutral bremsstrahlung, NBrS) in noble liquids~\cite{Borisova22}, provided that the momentum-transfer cross section is used in calculation of the NBrS cross section (instead of the energy-transfer one). Apart from the obvious achievement in discovering a new physical effect, the last statement puts an end to the long dispute about which cross section should be used in the NBrS formula. In particular, our results favour NBrS cross section approximation derived in~\cite{Milstein22,Park00}.

Despite the relatively low light yield, the effect of NBrS EL in noble liquids may pave the ways for new readout schemes in single-phase noble liquid detectors for dark matter searches and low-energy neutrino detection.


\begin{acknowledgments}
This work was supported in part by Russian Science Foundation (project no. 20-12-00008, \url{https://rscf.ru/project/20-12-00008/}).
\end{acknowledgments}

\bibliography{Manuscript}

\end{document}